%% file: _main.tex
\documentclass[runningheads]{llncs}
\usepackage[a4paper, total={140mm, 220mm}]{geometry}
\usepackage[T1]{fontenc}

\usepackage{comment}
\usepackage{graphicx}
\usepackage{hyperref}
\graphicspath{ {figures/} }
\usepackage{subcaption}
\usepackage{balance}
\usepackage{float}
\restylefloat{figure}
\usepackage{orcidlink} 

\usepackage{xcolor} % Ensure xcolor package is loaded

\newcommand\vldbdoi{XX.XX/XXX.XX}

\newcommand\vldbavailabilityurl{https://github.com/dliyanage/YCSB-IVS}
\usepackage{xspace}
\newcommand{\mainrun}{\emph{main-run}\xspace}
\newcommand{\cleanrun}{\emph{clean-run}\xspace}
\newcommand{\avgrun}{\emph{average-baseline}\xspace}
\newcommand{\spreadrun}{\emph{spread-baseline}\xspace}

\newcommand{\mongodb}{Mongo\-DB\xspace}
\newcommand{\innodb}{MariaDB+\-InnoDB\xspace}
\newcommand{\rocksdb}{MariaDB+\-MyRocks\xspace}
\newcommand{\techreport}{\href{https://github.com/dliyanage/YCSB-IVS/blob/master/analysis/Technical Report/YCSB_IVS_Tech_Report.pdf}{\texttt{YCSB-IVS Tech Report}}}

\usepackage{balance}

\begin{document}
\title{A Benchmark for Databases with Varying Value Lengths} %[Experiment, Analysis, and Benchmark]}

%%
%% The "author" command and its associated commands are used to define the authors and their affiliations.
\author{Danushka Liyanage\inst{1}\orcidlink{0000-0003-0295-1996} \and
Shubham Pandey\inst{2}\orcidlink{0009-0002-2986-792X} \and
Joshua Goldstein\inst{3} \and
\\ Michael Cahill\inst{1}\orcidlink{0009-0000-5834-8278} \and
Akon Dey\inst{4} \and
Alan Fekete\inst{1}\orcidlink{0000-0003-3804-5450} \and
Uwe R\"ohm\inst{1}\orcidlink{0000-0003-3598-7119}}

\institute{
University of Sydney, Sydney, Australia \\
\email{\{danushka.liyanage, michael.cahill, alan.fekete, uwe.roehm\}@sydney.edu.au} \and
Independent Researcher, Austin, Texas, USA \\
\email{shubhambeethoven@gmail.com} \and
Texas A\&M University, College Station, Texas, USA \\
\email{jgoldstein345@gmail.com} \and
Predictable Labs, Inc., Foster City, California, USA \\
\email{akon.dey@gmail.com}
}

\authorrunning{Liyanage et al.}

\maketitle

%\author{Danushka Liyanage}
%\orcid{0000-0003-0295-1996}
%\affiliation{%
%  \institution{University of Sydney}
%  \city{Sydney}
%  \state{Australia}
%}
%\email{danushka.liyanage@sydney.edu.au}

%\author{Shubham Pandey}
%\orcid{0009-0002-2986-792X}
%\affiliation{%
% \institution{Independent Researcher}
% \city{Austin}
% \state{Texas}
%  \country{USA}
%}
%\email{shubhambeethoven@gmail.com}

%\author{Joshua Goldstein}
%\orcid{0000-0002-1825-0097}
%\affiliation{%
%  \institution{Texas A\&M University}
%  \city{College Station}
%  \country{USA}
%}
%\email{jgoldstein345@gmail.com}

%\author{Michael Cahill}
%\orcid{0009-0000-5834-8278}
%\affiliation{%
%  \institution{University of Sydney}
%  \city{Sydney}
%  \state{Australia}
%}
%\email{michael.cahill@sydney.edu.au}

%\author{Akon Dey}
%\orcid{0000-0001-5109-3700}
%\affiliation{%
%  \institution{Predictable Labs, Inc.}
%  \city{Foster City}
%  \country{USA}
%}
%\email{akon.dey@gmail.com}

%\author{Alan Fekete}
%\orcid{0000-0003-3804-5450}
%\affiliation{%
%  \institution{University of Sydney}
%  \city{Sydney}
%  \state{Australia}
%}
%\email{alan.fekete@sydney.edu.au}

%\author{Uwe R\"ohm}
%\orcid{0000-0003-3598-7119}
%\affiliation{%
%  \institution{University of Sydney}
%  \city{Sydney}
%  \state{Australia}
%}
%\email{uwe.roehm@sydney.edu.au}

\begin{abstract}
The performance of database management systems (DBMS) is traditionally evaluated using benchmarks that focus on workloads with (almost) fixed record lengths. However, some real-world workloads in key/value stores, document databases, and graph databases exhibit significant variability in value lengths, which can lead to performance anomalies, particularly when popular records grow disproportionately large. Existing benchmarks fail to account for this variability, leaving an important aspect of DBMS behavior underexplored.

In this paper, we address this gap by extending the Yahoo! Cloud Serving Benchmark (YCSB) to include an ``extend'' operation, which appends data to record fields, simulating the growth of values over time. Using this modified benchmark, we have measured the performance of three popular DBMS backends: MongoDB, MariaDB with the InnoDB storage engine, and MariaDB with the MyRocks storage engine. Our experiments alternate between extending values and executing query workloads, revealing significant performance differences driven by storage engine design and their handling of variable-sized values.

Our key contribution is the introduction of a novel benchmarking approach to evaluate the impact of growing value sizes and isolate the effect of querying data with a distribution of data sizes from any cost associated with accessing data after a history of updates. 
%, a comparative performance analysis of popular DBMS systems under this workload, and insights into the performance challenges posed by large, non-uniform values. 
This highlights the need for more representative benchmarks that capture the dynamic nature of real-world workloads, providing valuable guidance for both practitioners and researchers.

\end{abstract}

%\pagestyle{\vldbpagestyle}
%\begingroup\small\noindent\raggedright\textbf{PVLDB Reference Format:}\\
%\vldbauthors. \vldbtitle. PVLDB, \vldbvolume(\vldbissue): \vldbpages, \vldbyear.\\
%\href{https://doi.org/\vldbdoi}{doi:\vldbdoi}
%\endgroup
%\begingroup
%\renewcommand\thefootnote{}\footnote{\noindent
%This work is licensed under the Creative Commons BY-NC-ND 4.0 International License. Visit \url{https://creativecommons.org/licenses/by-nc-nd/4.0/} to view a copy of this license. For any use beyond those covered by this license, obtain permission by emailing \href{mailto:info@vldb.org}{info@vldb.org}. Copyright is held by the owner/author(s). Publication rights licensed to the VLDB Endowment. \\
%\raggedright Proceedings of the VLDB Endowment, Vol. \vldbvolume, No. \vldbissue\ %
%ISSN 2150-8097. \\
%\href{https://doi.org/\vldbdoi}{doi:\vldbdoi} \\
%}\addtocounter{footnote}{-1}\endgroup

%\ifdefempty{\vldbavailabilityurl}{}{
%\addvspace{.3cm}
%\begingroup\small\noindent\raggedright\textbf{PVLDB Artifact Availability:}\\
%The source code, data, and/or other artifacts have been made available at \url{\vldbavailabilityurl}.
%\endgroup

\input{1-intro}
\input{2-motivation}

\input{3-related}
\input{4-benchmark}
\input{5-results}
\input{8-conclusion}

%\begin{acks}
% Acknowledgements come here.
%\end{acks}

\vspace{0.5cm}

\noindent\textbf{Acknowledgments.}
We thank the anonymous reviewers of TPCTC~2025 for their valuable feedback on the original version of this paper. This research was funded by Australian Research Council Discovery Project DP210101984.

%\clearpage
\balance
\bibliographystyle{splncs04}
\bibliography{references}

\end{document}

%% file: 1-intro.tex
\section{Introduction}

%importance and impact of benchmarks
%Performance evaluation of database management systems (DBMS) is a critical aid to understanding their behavior under various workloads, and the development of widely adopted benchmarks has been seen as a driver for improvements in DBMS implementation. Benchmarks choose data and workload characteristics that simplify real-world applications, but they aim to keep essential features that stress possible inefficiencies of implementation, for example, by having significant contention, requirements for sometimes rolling back, expectations to scale the amount of data along with the workload, etc. If some aspect of the DBMS is not exercised by benchmarks, our community's understanding of those issues may be weakened, and there is also a risk that platform engineers may not pay proper attention to designing for this. 

%limitation of existing benchmarks: little variation in record sizes
Benchmarks, such as the TPC suite, have served the database field well; they have helped practitioners with system and configuration choice, and have been a spur for researchers to improve platform design. 
In this paper, we consider a gap in prior work on benchmarking: the neglect of cases where some data items (records in tables, documents in a document store, values in a key value store) get longer and longer as the system evolves, eventually becoming quite long (say megabyte size). This can happen particularly when an item has an attribute whose type is array and where the application uses this array to store information about the item's history. In that case, some operations of the application append extra information to that attribute; if the item is a popular one and the database lasts a long time, the item may grow very long. This situation has not been well covered in previous benchmarks. Some (e.g., YCSB) have fixed size for all items; while TPC-C does include some variation in record size (due to VARCHAR fields such as C\_DATA), this variation is limited, with no record longer than 1KB. 

%impact of growth in item size
The existence in real applications of items that grow in size over time can impact the performance of the system, depending on choices made in the implementation of the storage layer. These situations can cause fragmentation of the data on disk or in memory, or may lead to frequent re-arrangements if the storage uses an update-in-place approach. Some systems store deltas to optimize the case where only a small part of an item is modified in an update. If these design aspects are not exercised in a benchmark, we will not know how well the choices work in practice. Thus, the focus of this paper is to propose a benchmark that examines what happens when items have lengths that increase significantly over time.

%need to disentangle three related changes
When the data includes some items that are growing steadily, there are several consequential changes to the data, and we aim to disentangle the performance impact of these. One impact of some records growing longer through history is that at any given time, the data items in the database will vary in length among themselves. Another consequence is that the total volume of data in the database will grow.  YCSB-IVS is designed to measure performance with a database where some items keep growing, but we also compare this with performance when the item sizes vary, but not as a result of a history of growth. Also, we look at the performance of two baselines with the same total volume and items of uniform size: one baseline has many small records, and the other has fewer records, which are all equally long.

%Traditional benchmarks for DBMS, such as TPC-C and TPC-H, often focus on relational database workloads where schemas define fixed row sizes, and values within these rows typically maintain constant sizes. Although this approach provides consistency for comparison, it fails to capture the variability and complexity of real-world workloads observed in key/value stores, document databases, and graph databases. In these systems, value sizes can vary significantly, influenced by application-specific usage patterns and data access behaviors.

%A particularly important example arises from ``popular'' records that grow disproportionately large due to frequent updates or attached data. Such records can cause pathological performance problems if the underlying DBMS is not equipped to handle them efficiently. Despite the practical relevance of these scenarios, existing benchmarks and performance studies rarely account for varying value sizes or their distributions. This gap in the literature motivates our study, which seeks to explore the impact of non-uniform value sizes on DBMS performance.

%overview of our approach
We propose a benchmark called YCSB-IVS which builds on the framework provided by Yahoo! Cloud Serving Benchmark (YCSB), a widely used benchmarking tool originally for key/value stores, but also with bindings for a wide variety of platforms including SQL ones. While the original YCSB workload has fixed size items (by default, 1000 bytes exactly), which can be read, updated, etc., we introduce an ``extend'' operation that appends data to one or more fields of a record, causing the length to grow over time. In contrast to the original YCSB, where data is loaded initially, and then a single measured run executes a mix of read, update, etc. operations, YCSB-IVS loads data, and then proceeds in a sequence of epochs, each of which performs some number of extend operations applied to keys randomly chosen according to some distribution, and then we measure the metrics (throughput, latency, etc.) for standard YCSB workloads against the state at the end of the extension phase.  As we run successive epochs, some of the data items become longer and longer, and the standard measurements show the slowdown that this causes. For comparison, we also measure for each epoch the read performance on a database with the same logical state (and thus the same distribution of item lengths) but no history of increase. We report the performance metrics for each epoch, and also report on two baselines which execute the queries against two datasets with equal-sized items and the same total volume as the given epoch; one baseline has the same number of items as in our benchmark (and so each item has length, which is the average length found after extensions). In the other baseline, all records have the unextended length, and so there are many more items than in the benchmark execution.  

We have applied YCSB-IVS to measure the performance of several popular DBMS platforms -- \mongodb, MariaDB with the InnoDB storage engine, and MariaDB with the MyRocks storage engine. Here we show a few examples of the results; full details are in our \techreport.

Our findings reveal significant performance differences between systems, driven by their storage engines and handling of large, variable-sized values. For example, some systems demonstrate robust performance even as values grow, while others exhibit surprising slowdowns. These results underscore the importance of incorporating varying value sizes into benchmarking methodologies to uncover hidden performance characteristics and better inform DBMS design and optimization.

%This paper makes the following contributions:

\begin{comment}
\begin{itemize}
\item \textbf{A novel benchmarking approach for databases/key-value stores:} We extend YCSB to simulate realistic scenarios where value sizes grow over time, reflecting issues that occur in real-world workload patterns, but have not been effectively benchmarked before.
\item \textbf{Extensive evaluation of popular DBMS platforms under YCSB-IVS:} We present detailed performance analyses of \mongodb, MariaDB with InnoDB, and MariaDB with MyRocks, highlighting the unique strengths and weaknesses of each system in the face of items that grow in length through time.
\item \textbf{Insights into DBMS performance under variable value sizes:} Our study identifies key factors that influence performance in the presence of large and non-uniform values, offering guidance for practitioners and researchers.
\end{itemize}
\end{comment}

In Section~\ref{Section:Motivation} we discuss in more detail some real-world cases where items grow in length substantially over time. Section~\ref{Section:Related} provides a critical analysis of prior work relevant to this research domain. In Section~\ref{Section:Benchmark} we describe the detailed operation of our new benchmark. In Section~\ref{Section:Results} we illustrate its use by reporting some measurements.

Our key contribution is to propose a novel benchmarking approach for databases/key-value stores. We extend YCSB to simulate realistic scenarios where value sizes grow over time, reflecting issues that occur in real-world workload patterns but have not been effectively benchmarked before. Through this work, we aim to broaden the scope of DBMS performance evaluation, expanding from the current focus on fixed-size data items to include workloads where some items can grow to be quite large. In doing so, we hope to contribute to a deeper understanding of how modern DBMS systems perform under conditions that are more closely aligned with real-world applications. The source code of YCSB-IVS and data from our experiments are provided at \url{https://github.com/dliyanage/YCSB-IVS}

%% file: 2-motivation.tex
\section{Motivation}\label{Section:Motivation}

%MongoDB \cite{mongodb_manual} is a document DBMS in which documents consist of  key/value pairs and are usually represented by JSON.

In numerous applications, array-valued attributes are a common feature within data records, playing an essential role despite not fitting the strict criteria of the first normal form (1NF) in traditional relational database models. These array attributes enable rich representations of complex data relationships and historical changes within the records. Despite their departure from conventional normalization principles, many relational databases (RDBMS) and modern NoSQL databases, such as document stores and graph databases, offer support for storing such multi-valued fields.

In practical applications, these array-type fields often serve critical functions. For example, an e-commerce platform might maintain a product record, complete with an array field that lists all previous prices at which the product was marketed. Each time the product is listed with a new price, the list of prices within the array attribute expands with the addition of the latest offered price. Similarly, a social media site keeps track of posts and tags. Over time, additional tags are appended to the existing list of tags for each post.

In long-term database usage, array-valued fields in popular items can experience significant growth, reaching megabytes in length, while entries in less popular items may only reach kilobyte ranges. This stark divergence in item size within a single collection or table is characteristic of data that has grown incrementally over time. The potential for such wide variations in record size can have implications for database performance. It underscores the need to understand and benchmark how the gradual expansion of array fields influences the overall performance of the DBMS.

We describe real-world situations from the industrial experience of some authors.

%In some workloads, applications push new entries into arrays. This causes documents to grow in size, which can affect performance by changing cache behavior and fragmenting memory. Over time, growing value sizes unevenly leads to a highly non-uniform distribution of value sizes. Reading very large values can be significantly slower than reading small values, and it is difficult to predict performance when mixing value sizes in a single workload.  

%Previous benchmarks have not extensively explored the behavior of databases system with non-uniform and growing value sizes.

%In this work, we explore the performance of workloads as value sizes grow, including their impact on query latency, throughput, and storage efficiency. By examining how three popular DBMS backends handle these scenarios, we aim to uncover important insights into their performance characteristics. Our study provides a deeper understanding of how variable value sizes interact with storage engines and highlights areas for potential optimization. This motivates the need for improved benchmarking methodologies to evaluate database systems under conditions that mirror real-world workloads more closely.

%\todo{dgraph experience with value growth}
%\subsubsection*{\textbf{Record growth in Dgraph}}
{\textbf{Record growth in Dgraph.}}
%Dgraph \cite{dgraph} is a platform that provides direct support for a data model as a graph. It is targeted at distributed systems of large scale with high update rates.
Dgraph is a distributed graph database platform optimized for high-throughput workloads. It stores graph-structured data by decomposing it into individual facts, called predicates. Each fact represented as a key-value pair is stored in Badger \footnote{\url{https://github.com/hypermodeinc/badger}}, an LSM-tree-based key-value store. Alongside this data, Dgraph also stores metadata, including type definitions and lists of unique identifiers (\emph{uids}) for each type \cite{jain2021dgraph}.
%Data storage patterns in Dgraph can lead to a disproportionate growth of values associated with a small subset of keys. This issue became particularly evident at Dgraph Labs, where it contributed to extended support calls, production outages at high-profile customer sites, and numerous sleepless nights.
%Dgraph relies on the Badger key-value store for data persistence. 
As the database matures and contains more data, the length of the values for a small number of keys, such as the types, scales in proportion to the overall growth of the application's data. This issue became particularly evident at Dgraph Labs, where it contributed to extended support calls, production outages at high-profile customer sites, and numerous sleepless nights.

%former version
%Some keys, related to type information, gradually accumulate a large number of identifiers (\emph{uids}) referencing records of that particular type \cite{jain2021dgraph}. As the database matures and contains more data, the size of these values scales in proportion to the overall growth of the database.

Compounding the issue, record deletions provide little relief due to the underlying storage mechanisms. Even with garbage collection in place, large record sizes remain a persistent challenge, exacerbating storage inefficiencies and operational complexity over time.

%\subsubsection*{\textbf{Document growth in \mongodb}} 
{\textbf{Document growth in \mongodb}.}
\mongodb is a platform that directly supports a document data model. Certain application patterns with \mongodb can lead to documents expanding over time, such as maintaining an exhaustive history of events or tracking changes to a value across multiple iterations. In these scenarios, what begins as small increments to a document's size can compound over time, leading to substantial increases in document size after numerous updates.

This growth pattern is particularly concerning when each update causes the document to swell incrementally; without careful design, such expansion can precipitate memory and storage fragmentation. \mongodb addresses this challenge by delaying when small updates to large documents are persisted to storage \footnote{\url{https://www.mongodb.com/community/forums/t/in-place-partial-updates/240487}}. Instead, they are cached and consolidated into fewer and larger updates upon subsequent write operations. This strategy mitigates the quadratic increase in work associated with handling numerous tiny updates and reduces the incidence of fragmentation.

The impact of this design choice is twofold: first, it alleviates the memory and storage overhead that would otherwise result from frequent small updates. Second, certain workloads experience considerable performance enhancement due to this optimization.

%% file: 3-related.tex
\section{Related Work}\label{Section:Related}

%\subsubsection*{\textbf{Databases Value/Object Sizes.}}
{\textbf{Databases Value/Object Sizes.}}
The size of values or objects in databases plays a crucial role in query execution, as larger sizes impose greater processing demands, thereby slowing down performance. Consequently, object size has been a central consideration in numerous query optimization cost models \cite{du1992query,gardarin1996calibrating}, often serving as a critical factor in addressing scalability challenges. For instance, Shapiro et al. \cite{shapiro1999managing} evaluated the performance of databases utilizing Binary Large Objects (BLOBs) and observed that increases in both database size and average object size significantly degrade transaction efficiency. %highlighting a key limitation in database scalability. 

It is evident that database developers have long recognized the trade-offs associated with value sizes, particularly the efficiency gains achieved by keeping value sizes small. This understanding has made data \emph{compression} a fundamental feature in modern databases and key-value stores \cite{iyer1994data,westmann2000implementation,idreos2020key}. A quantitative assessment by Hurson et al. \cite{hurson1993object} in the early 1990s demonstrated that smaller object sizes significantly improve the efficiency of object clustering, reducing I/O operations and achieving cost improvements ranging from a factor of 2 to 15.

Although efforts are often made to keep value or object sizes as small as possible, there are numerous scenarios in which value sizes naturally grow over time. Pelekis et al. \cite{pelekis2004literature} highlighted that size plays a critical role in spatio-temporal database models, especially in databases that undergo continuous changes.

%\subsubsection*{\textbf{Database Benchmarking.}}
{\textbf{Database Benchmarking.}}
The `Wisconsin' benchmarks \cite{bitton1983benchmarking}, introduced in the early 1980s, were among the first systematic attempts to benchmark relational database systems, evaluating platforms like DIRECT, INGRES, ORACLE, and IDM500. Soon after, the TP1 benchmark (1985) emerged, measuring transaction throughput through components such as \emph{transaction (DebitCredit)}, \emph{OS and I/O (Sort)}, and \emph{file system (Copy)}. Jim Gray's benchmarking handbook \cite{gray1993database} formalizes database benchmarking practices, leading to the expansion of benchmarks from the Transaction Processing Performance Council (TPC) \cite{TPC}.  

The TPC benchmarks became influential, and \emph{DebitCredit} evolved into TPC-A and then TPC-C for OLTP \footnote{OLTP - Online Transaction Processing.} workloads. To address both OLTP and OLAP \footnote{OLAP - Online Analytical Processing.} systems, the hybrid TPC-CH benchmark was introduced \cite{funke2011benchmarking}, bridging TPC-C (OLTP) and TPC-H (OLAP). Building on TPC-H, the \emph{Star Schema Benchmark (SSB)} was developed to measure database performance in traditional warehousing applications \cite{o2009star}.  

Over time, specialized benchmarking tools emerged to evaluate diverse database architectures. Examples include a variety of benchmarks inspired by social network graph structures including LinkBench \cite{armstrong2013linkbench} and TAOBench ~\cite{cheng2022taobench}, \emph{Jackpine} for spatial databases \cite{ray2011jackpine}, SmartBench \cite{gupta2020smartbench} for IoT support. LDBC is a consortium which has defined a family of benchmarks for graph data, eg \cite{ErlingALCGPPB15}. YCSB was introduced to measure performance of cloud-based key-value platforms \cite{ycsb}, and because it offers an extensible framework, later work used YCSB as the basis to incorporate other aspects including transactions \cite{DeyFNR14}, geospatial data \cite{KimHYK23}.
Recently, \cite{BensonBBLLPRSST24} focused on the danger of systems becoming tuned for widely-used benchmarks, and advocates for introducing new ``surprise'' aspects each year.

%\subsubsection*{\textbf{Value Size in Benchmarking.}}  
{\textbf{Value Size in Benchmarking.}}  
Large object sizes have been considered in database benchmarking due to their direct impact on overall database size. Biliris \cite{biliris1992performance} analyzed the performance of large object management across three database systems from 1992—EXODUS \cite{Carey1986}, Starburst, and EOS—focusing on length-changing updates. The study proposed strategies for segment size adjustments to enhance operational efficiency while noting the trade-off between performance gains and reduced storage utilization with large fixed-size segments.  

In past decades, the main relational database benchmarks have predominantly employed fixed (or only slightly varying) item sizes in their setups \cite{battle2020database}. While scaling the total database volume has been a traditional benchmarking consideration (e.g. ``scale factor'' in the TPC benchmarks), the dynamic nature of individual value sizes and their distribution over time have received less attention. 

Cao et al. demonstrated in RocksDB-based production systems such as UDB, ZippyDB, and UP2X at Facebook that the distribution of key and value sizes is closely tied to specific use cases and applications \cite{cao2020characterizing}. A recent study considered space-optimising storage structures in Neo4j under real application loads, and found little impact on overall performance, ascribed perhaps to the limited importance of indexing in that platform \cite{TheodorakisCW24}.

%% file: 4-benchmark.tex
\section{YCSB-IVS Benchmark Description and Design}
\label{Section:Benchmark}

The main contribution of this paper is a new benchmark called YCSB-IVS for \emph{YCSB - Increasing Value Sizes},  which can assess the impact on database performance as the size of data items grows over time. We work within the framework of Yahoo! Cloud Serving Benchmark (YCSB\footnote{YCSB: \url{https://github.com/brianfrankcooper/YCSB.git}}) \cite{ycsb}. This is a widely used framework to evaluate the performance of NoSQL and SQL databases and other cloud-based data delivery systems. Designed to facilitate comparative analysis, YCSB provides a set of core workloads that simulate varied real-world database use cases, including read-heavy, write-heavy, and balanced scenarios. The workloads are mixes of Read, Update, Insert, Delete and Scan operations, with configuration parameters that control the distribution of the keys involved as well as the proportion of the operation types. The framework enables us to measure crucial performance metrics, such as throughput and latency (average and tail), under different configurations. YCSB’s extensibility allows for customisation of workloads, database bindings, and distributions, making it a versatile tool for assessing various aspects of data serving systems under controlled conditions.
As YCSB is a benchmark that assumes only a key-value logical data model, we implement a \emph{value-length extension} functionality, which increases the length of the value associated with a key; due to the generality of this logical data model, we can apply the benchmark to platforms with richer data models, including relations or JSON documents.

%\subsubsection*{\textbf{Extend Operation.}}
{\textbf{Extend Operation.}}
 To see the impact of growing value sizes on database performance, we implemented an extra operation, which will increase the length of value for a key chosen according to a specific distribution; this is similar to the original Update of YCSB, in that a key is chosen from a distribution, and the associated value is replaced by a new one; unlike the provided Update, our Extend associates to this key a new value which is longer than the previous one by the amount specified in the workload configuration as the \texttt{extendfieldlength} parameter (we have a default increase of 100 bytes). 
 %check this please
 Since YCSB data model is multiple (default ten) string-valued fields as the value for each key, each extend operation randomly chooses one field uniformly and replaces the contents of that field by a longer string. 
 
 %Although this new method could readily reduce value sizes, we opted against this approach, as reductions could offset increases or introduce randomness into the overall size distribution.

%To enable and control the number of data extension operations, we introduced an additional configuration parameter, \texttt{extendproportion}, alongside the other operation-related proportions (e.g. \texttt{updateproportion}). A nonzero \texttt{ expandproportion} results in value extension operations proportional to that value, relative to the total \texttt{operationcount}. This integration of the value extension functionality allows us to run workloads with mixed operations incorporating both value extensions and other operations, as well as phases consisting of 100\% value extension operations. We call our proposed workload YCSB-IVS (\emph{YCSB - Increasing Value Sizes}).

%\todo{Add a description of the interaction between extends and requestdistribution to create different (static) distributions of value size}

%Recalling that the \emph{extend} operation is functionally analogous to the \emph{update} operation, value size extensions occur based on the specified \emph{requestdistribution}. For example, under a uniform distribution, keys are selected uniformly at random for extension, whereas a Zipfian distribution prioritises extending the values of frequently accessed (popular) keys located at the head of the distribution \cite{ycsb}. 

\subsection{Benchmark Methodology}
\label{Section:Benchmark-Methodology}

The original YCSB framework begins with a {\em load} phase where data is generated and loaded to the data store being evaluated, and then performance is measured during a {\em run} phase, where simulated clients submit operations (such as Read, Update, etc.) against the data store. In the original framework default configuration, the data is a collection of items all exactly 1000 bytes long. In our YCSB-IVS, we have a more complex experimental approach to capture performance when data items can grow; the initial data load is followed by a succession of {\em epochs}; each epoch has two phases, first an {\em extend} phase in which the Extend operation is performed many times, thus growing the length of some of the records, and then the epoch has a {\em run} phase where we measure throughput and latency for a mix of operations, against the data that resulted from the extend phase of this epoch. Then the next epoch begins, with further extending record lengths and then another measured run phase. In fact, we use the YCSB framework's run capability for both our extend and run phases in each epoch: our extend phase is simply a YCSB run of a workload containing 100\% Extend operations, and our run phase can be an arbitrary mix of operations specified in the workload configuration. We make sure that the measurements we capture in our run phase of a given epoch are all against a data set with a particular distribution of item lengths, which grows from epoch to epoch due to the extend phases. Our experiment results display the way the metrics (throughput and latency) change as the epochs progress.

\begin{figure}[tbh]
    \centering
    \includegraphics[width=\textwidth]{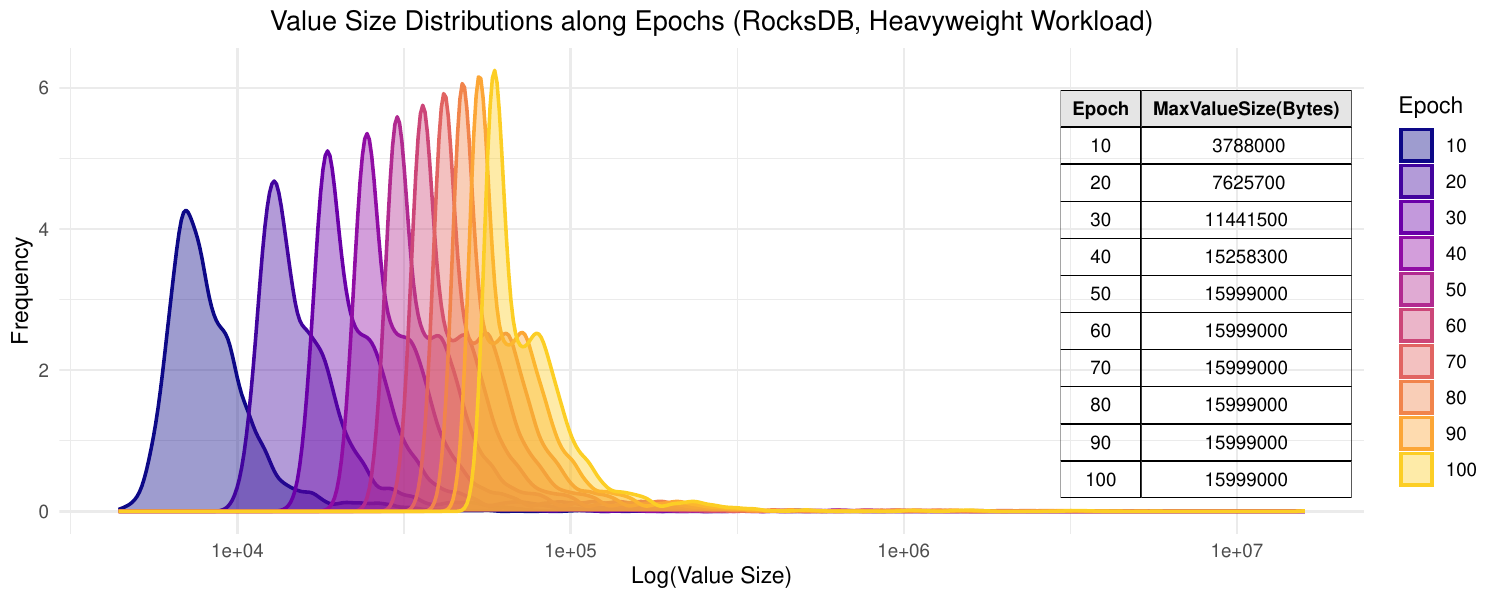}

    \caption{Record size distribution and maximum value size change along the epochs with Zipfian distribution of extends.}
    \label{fig:innodb-heavy-zipf-size-dist}
\end{figure}

%\subsubsection*{\textbf{Maximum record size}}
{\textbf{Maximum record size}.}
There is a pragmatic issue that arises from restrictions in the platforms with which we work and leads to some subtlety in the way we configure YCSB-IVS. 
Many database systems impose restrictions on the maximum size of individual records, stemming from both design considerations and practical limitations. For example, \mongodb restricts documents to a maximum size of 16MB, as detailed in their official documentation\footnote{\url{https://www.mongodb.com/docs/manual/reference/limits/\#bson-documents}}. MariaDB, while more flexible, adheres to its own constraints: the LONGTEXT data type must be used for fields exceeding 64KB\footnote{\url{https://mariadb.com/kb/en/text/}}. However,  with MariaDB's default settings, a client application can create a row larger than 16MB, yet that row cannot be successfully read due to an underlying limit on the packet size.

Given these constraints, we designed YCSB-IVS to incorporate a cap on record sizes by limiting the size of individual fields. When an "extend" operation encounters a case that, if extended, would exceed its maximum allowable size, it is bypassed entirely, and no update is applied. In YCSB-IVS, the default field size limit is set at 1.6MB, which, in consequence, sets a record's upper boundary on length at just below the 16MB threshold to circumvent these platform-specific limitations. This choice affords us several advantages: it provides a workable solution that sidesteps platform constraints while having minimal impact on the workload's execution.

Our experiments, with the Zipfian request distribution for extend operations, target a small set of records to be extended most often. As shown in \autoref{fig:innodb-heavy-zipf-size-dist}, the Zipfian distribution of extend operations both grows the typical record length \emph{and} makes record lengths increasingly skewed as the epochs advance.
Yet, we 
%can see from \autoref{fig:value-extension-costs} 
found that no more than two records reached this maximum limit. This validates our design decision, demonstrating that even under stringent conditions, the limit is sufficient to prevent record expansion beyond what the database systems can handle while minimizing the impact of the limit on ordinary operation of the workload.  This limitation serves as a protective measure, ensuring the robustness and reliability of YCSB-IVS against the backdrop of real-world database constraints.

%\subsubsection*{\textbf{Extend Phase.}}
{\textbf{Extend Phase.}}
This phase involves submitting Extend operations against the database, with keys chosen according to the \texttt{requestdistribution} parameter which is specified in the configuration. In our experiments, we consider both Uniform and Zipfian distributions for the keys to be extended (these distributions are in the default YCSB codebase). For the Zipfian distribution, we use the YCSB default parameter 0.99. Extending keys chosen as a Zipfian distribution means that a few data items will be extended much more frequently than others, and these popular items will therefore grow quite long as the epochs pass. An item which is chosen in an extend operation grows in length by the amount given as the \texttt{extendfieldlength} configuration parameter, defaulting to 100 bytes (this default is used for all the experiments we report here). We also use the data platform's backup capability to take a complete dump of the logical data state at the end of each {\em extend phase}, for use in the \cleanrun we describe later.

{\textbf{Run Phase.}}
%The {\em run} phase of an epoch in YCSB-IVS has 100\% Read operations (no Scan, Insert, Update or Delete operations). 
A workload comprising any arbitrary mix of Read, Scan, Insert, Update, or Delete operations can be performed within the {\em run phase} of an epoch in YCSB-IVS. Workloads involving Insert and Update operations may cause the value size distribution to change during the {\em run phase}. To mitigate this deviation, we configure YCSB to use a value size histogram that is calculated at the beginning of the run phase with a bin size of 100 by setting the parameter \texttt{fieldlengthdistribution=histogram}.

The number of operations in an epoch is controlled by the \texttt{operationcount} configuration parameter and is chosen to be large enough so that any startup effects from running the YCSB client are amortized. 
%In our testing, 10,000 operations per epoch were insufficient and startup effects were observed, noting that YCSB does not support a warm-up phase before measurement begins. 
The experiments we report here use 100,000 operations per run phase. For this phase, the keys are chosen by the uniform distribution regardless of which distribution is used for the extend phase. Using a uniform distribution means that operations during a workload will be evenly distributed across value sizes. This ensures that the logical work done during the run phase depends on the total data volume but not the distribution of value sizes. In each epoch's run phase, we use the YCSB facility to capture the throughput and latency metrics (average or 99th percentile).

%We highlight that maintaining a static value size distribution during performance measurement is crucial for reliable and interpretable results. After each \emph{extend} phase, the value size distribution differs from the previous round due to the extension of values. During the subsequent \emph{run} phase, database operations consisting of reads, updates, scans, inserts, and deletes are executed to assess performance. While \emph{read} and \emph{scan} operations leave individual values unmodified, operations such as \emph{insert}, \emph{update}, and \emph{delete} inherently alter the value size distribution.

%To ensure the interpretability of performance metrics, it is essential to isolate the impact of value size changes on the database's behavior. In this paper, we exclusively focus on read-only workloads, enabling us to systematically evaluate the influence of evolving value size distributions on database performance over time without confounding effects introduced by operations that modify the data.

%\todo{Explain that we can't mix extend operations with existing YCSB workloads in a simple way because updates and inserts don't respect the value size distribution}

%\todo{Mode: describe run, dump/load, \avgrun}
%\subsubsection*{\textbf{Experiment Modes.}}
{\textbf{Experiment Modes.}}
As described above, the \textbf{\mainrun} epochs in our experimental methodology involve iteratively executing \emph{extend} and \emph{run} phases following the initial data loading, to show how performance changes as some data items grow in length over time due to Extend operations. This is a simplified version of what we have seen in real-world scenarios. However, one might consider whether the performance is based on the history of growth in some records over time, or instead is just a consequence of having records of different sizes in the database state, without any change through time. So, for better understanding, we have included in our benchmark methodology another set of epoch-based measurements that aim to disentangle these issues. 
%As well as the history of repeated growth (which can cause fragmentation, cache problems, long version chains, etc), the data also develops a wide variation in item lengths at each given epoch. 

%We call an extend phase followed by a run phase an \emph{epoch}, and the workload consisting of a sequence of epochs the \mainrun workload.

%Our methodology calls for also performing another set of epoch-by-epoch experiments to disentangle whether performance changes are due to history of growth or non-uniformity of length.
In YCSB-IVS, we measure the \textbf{\cleanrun} mode, in which, in each epoch, the logical data backup taken at the end of the extend phase of \mainrun is restored to a fresh instance of the data platform, and then the run operations are submitted to this clean copy of the data.
That is, in any epoch of \cleanrun, we have a database state which is logically the same as that in the corresponding epoch of \mainrun, but without any history of change over time, so the physical structures will likely be simpler. Any difference between \mainrun and \cleanrun reveals an effect that is due specifically to change in record lengths through an execution.

The baselines use only prior YCSB functionality (that is, there are no Extend operations, and all records have uniform length). They are measured separately, rather than after each extend phase, to show the impact of simple impacts of the scale of data volume, separate from non-uniformity. We perform the baselines for a range of scales corresponding to the data volume seen in the dump at the end of an epoch (this can in fact be calculated from the number of extend operations performed in the epochs up until then and the extra length added in each operation). \textbf{\spreadrun} measures the effect of having many items of a fixed size, and \textbf{\avgrun} measures the effect of a fixed set of long data items. In \avgrun, the data volume is distributed equally among the same number of records as in \mainrun; this ensures uniform item lengths which are quite long, and longer at increasing scale. The data with this uniform length is loaded into a fresh instance of the platform, and the read-only operations are run against that. In contrast \spreadrun for a given data volume has all records with the same length as those in the initial load of \mainrun, before any extensions; and thus the scale in data volume is achieved by having a very large number of records in \spreadrun (far more than in \mainrun). 

As expected (and mostly confirmed in our measurements), the performance of \spreadrun is mostly independent of the scale of the data volume; the indexing in each platform is effective and so the reads are not much affected when there are many more records to choose from. In contrast, each platform shows substantial slowdown with long records, though the details vary. This is expected since each read operation accesses and outputs the whole record. In our charts, we will show the metrics for \mainrun and \cleanrun at each epoch, and also for comparison the metrics from \avgrun at the appropriate data volume.

Any substantial difference between the metrics in \cleanrun for an epoch, and \avgrun for the data volume at the end of that epoch, indicates that the platform is affected by variations in item lengths. In contrast, any difference between \cleanrun and \mainrun can be attributed specifically to the gradual growth of item lengths over time and its impact on physical storage structures. If all three modes yield similar metrics, then the determining factor is simply the total record length, rather than variations in individual item sizes.

Each result chart shows curves for a performance metric against the epoch number.

%To comprehensively evaluate the impact of value size extensions under varying conditions such as request distribution and data/operation volumes, we conduct two complementary experimental modes alongside the mainrun.  

%\begin{itemize}
 %   \item \cleanrun: After each \emph{extend} phase, the database is backed up and restored (i.e., dumped and reloaded) to create a fresh instance identical to that of the \mainrun. This approach aims to isolate and evaluate performance discrepancies that arise from runtime value size extensions, while eliminating confounding factors such as changes in data structure layout and caching effects.
    
 %   \item \avgrun: Following each \emph{extend} phase, the total size of all values across the key range is redistributed evenly, ensuring that each key has values of equal size. This mode allows us to isolate the impact of performance variations caused by the spread of value sizes. By comparing this deterministic distribution with the stochastic distributions in other modes, the \avgrun highlights differences in performance attributable to variability in value sizes.
%\end{itemize}

%\subsubsection*{\textbf{Performance Evaluation Matrices.}}
{\textbf{Performance Evaluation Matrices.}}
We utilize \emph{latency} and \emph{throughput} as the primary metrics for evaluating system performance, reflecting their widespread adoption in performance analysis \cite{wang2008measurement}. The YCSB framework provides overall throughput, and several measures of the distribution of latencies for each type of operation (Read, Update, etc). We capture them during both the \emph{extend} and \emph{run} phases of our experiments, though our charts here focus on the measurements of the run phases, and we mostly report throughput, average latency for each operation type, and 99-percentile latency for each operation type. We often perform several (say, five) complete runs of the benchmark, and show the average value for a metric, and also a spread (from the mean to plus/minus 1.96 time standard error among the measured runs, all at the given epoch).

\subsection{Scale of Benchmark} \label{section:setup}

In \autoref{tab:experiment-configuration} we describe the default settings for YCSB-IVS. We use these settings for the results reported in \autoref{Section:Results}.

\begin{table}[ht!]
\centering
\caption{Experiment default configuration.}
\begin{tabular}{|l|r|}
\hline
\textbf{record count} & 10000 \\ \hline
\textbf{field count per record} & 10 \\ \hline
\textbf{initial field length} & 100 \\ \hline
\textbf{extend operations during extend phase, per epoch} & 100000 \\ \hline
\textbf{added length in one extend operation} & 100 \\ \hline
\textbf{operations in run phase, per epoch} & 100000 \\ \hline
\end{tabular}
\label{tab:experiment-configuration}
\end{table}

Since there are 10000 records, which start at 1000 bytes, and each extend operation adds 100 bytes to a
record, the database starts with about 10MB of data
and grows to approximately 1GB. While this volume of data isn't enough to stress the caching
or I/O of each system, our focus is on the cost to access records of varied
lengths in cached data. By the time we have completed 100 epochs, one experiment run has performed 10 million operations in its run phases.

We have also performed some experiments (not reported here) with a smaller-scale ``lightweight'' workload, with 1000 records, and 10,000 extend operations in each extend phase, and 10,000 other operations in each run phase. However, these runs were sensitive to startup effects.

%% file: 5-results.tex
\section{Case Studies of the Benchmark}
\label{Section:Results}

To illustrate the use of the benchmark, we present here a selection from the results when using  the benchmark with a range of platforms.
More of the results are discussed in \techreport. Data for all the experiments is at \url{https://github.com/dliyanage/YCSB-IVS}.

\subsection{\textbf{Experimental Setup}}
%We evaluate our proposed benchmark on multiple database platforms over multiple experimental trials.

%\subsubsection*{\textbf{Platforms.}} 
{\textbf{Platforms.}} 
%To evaluate the impact of changes in value size in a diverse range of database architectures, 
We apply the benchmark with three widely used database platforms. These DBMSs, summarized in Table~\ref{tab:databases}, represent a cross section of logical data models and physical storage structures. \mongodb and \rocksdb were run with default settings \cite{myrocks_vars,mongodb_vars}.  \innodb was run with the buffer pool size adjusted by setting \texttt{innodb\_buffer\_pool\_size} to 4GB\footnote{\url{https://mariadb.com/docs/server/storage-engines/innodb/operations/configure-buffer-pool/}}.

\begin{table}[htb]
\centering
\caption{Utilized DBMSs and their characteristics.}
\resizebox{0.75\columnwidth}{!}{%
\begin{tabular}{|c|c|c|c|c|}
\hline
 & \textbf{Database}       & \textbf{Version} & \textbf{Logical Model}          &  \textbf{Storage Structure} \\ \hline
1              & \rocksdb     & 10.6   & Relational                     & LSM-tree                 \\ \hline
2              & \innodb     & 10.6      & Relational                & B-tree                   \\ \hline
3              & \mongodb        & 6.0         & Document                       & B-tree                   \\ \hline
\end{tabular}%
}
\label{tab:databases}
\end{table}

%To demonstrate its utility, we present detailed results for \rocksdb, highlighting how the benchmark reveals performance variations as the size of the data items increases over time. The results for the other two database systems listed in Table~\ref{tab:databases} are available in the supplementary materials. 

%\subsubsection*{\textbf{Infrastructure.}}
{\textbf{Infrastructure.}}
The experiments were conducted on an AWS EC2 \texttt{m7i.xlarge} instance running Ubuntu 24.04 LTS. The machine is equipped with a dual-core Intel Xeon Platinum 8488C processor (Sapphire Rapids architecture) with a 105 MB L3 cache and hyper-threading enabled. It includes 16 GB of RAM and a 30 GB NVMe-backed Elastic Block Store for storage. Networking is handled by an Elastic Network Adapter, and the instance is configured without a graphical interface to optimize for computational workloads.

\subsection{Impact of Average Length of Values}

\begin{figure}[htb]
    \centering
    \includegraphics[width=0.7\columnwidth]{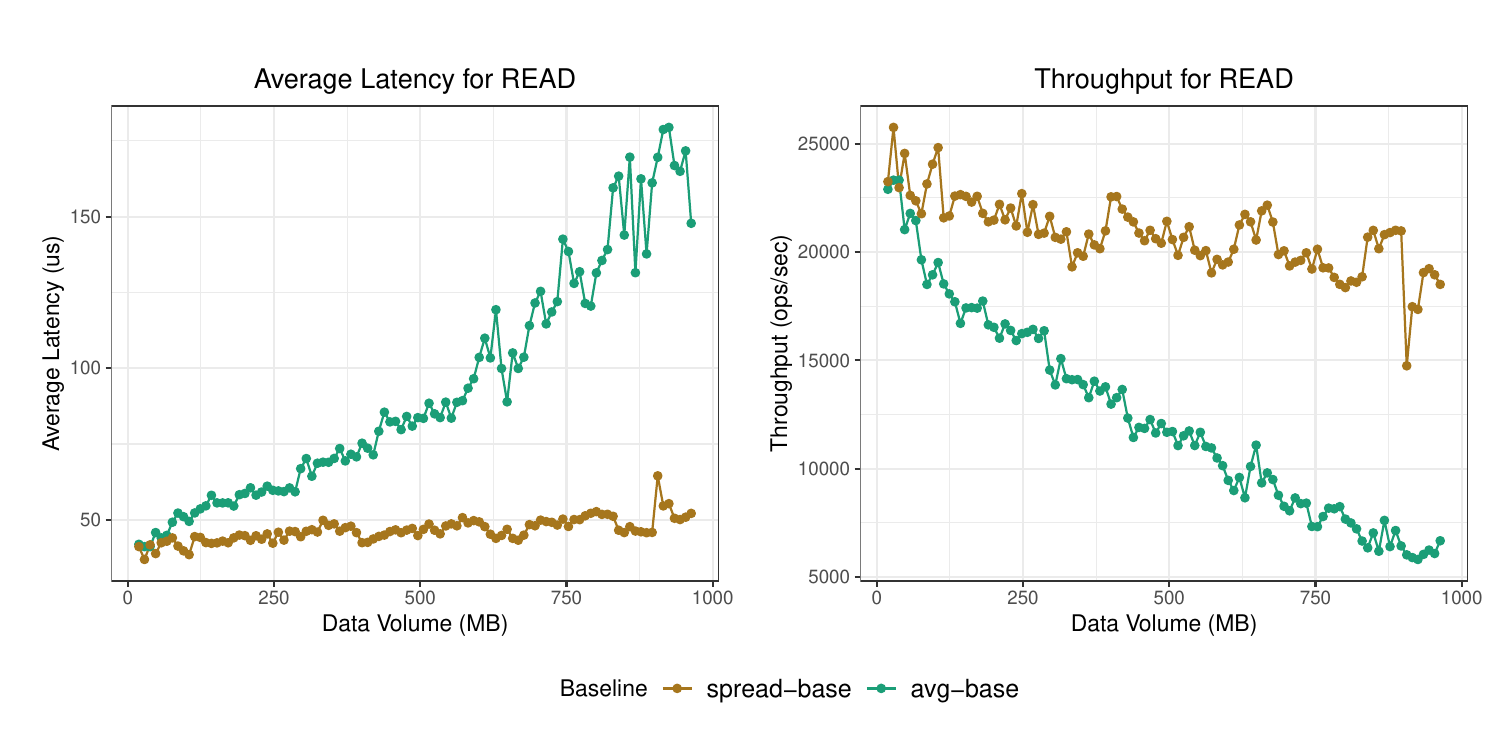}
%    \caption{Query performance baselines for \rocksdb, where a given data volume is created in \spreadrun by varying the number of fixed-size records, or in \avgrun by varying the size of a fixed number of records.}
    \caption{Baselines for \rocksdb, for YCSB Workload C.}
    \label{fig:myrocks-baselines}
\end{figure}

%\subsubsection*{\textbf{Baselines.}}
In \autoref{fig:myrocks-baselines}, we illustrate how one platform (\rocksdb) performs for the two baselines. Note that here, the data is not formed using Extend operations; rather for each data volume on the x-axis, we apply the YCSB Workload C directly to a database where all records are initialised to have equal length. One baseline, \spreadrun, involves obtaining the indicated total volume of data through as many records as needed, all of which have the same size (1KB) corresponding to the initial size in our benchmark runs. In contrast, \avgrun has the given data volume through the same  number of records as in the  benchmark (10,000), with whatever equal size is then needed for the given volume. Here the x-axis is data volume (which correlates linearly with the epoch numbers in other charts).

As expected, due to an effective index on the key used for reads, \spreadrun shows that \rocksdb experiences a slight gradual decline in read throughput, with some variance as the data volume scales through the presence of more records. However, the average read latency remains largely stable, suggesting that the addition of small records does not significantly impair \rocksdb's ability to serve read operations efficiently.

In contrast, under \avgrun, we observe a clear degradation in throughput as record sizes increase, along with significant variance in latency. This is expected, since records become longer, and each read operation transfers the whole record. 
%This performance decline highlights the sensitivity of \rocksdb to growing record sizes and underscores the importance of benchmarking workloads that vary data sizes to fully capture database behavior under realistic conditions.

% Placed here so page layout flows
\begin{figure}[thb]
    \centering
    \includegraphics[width=\columnwidth]{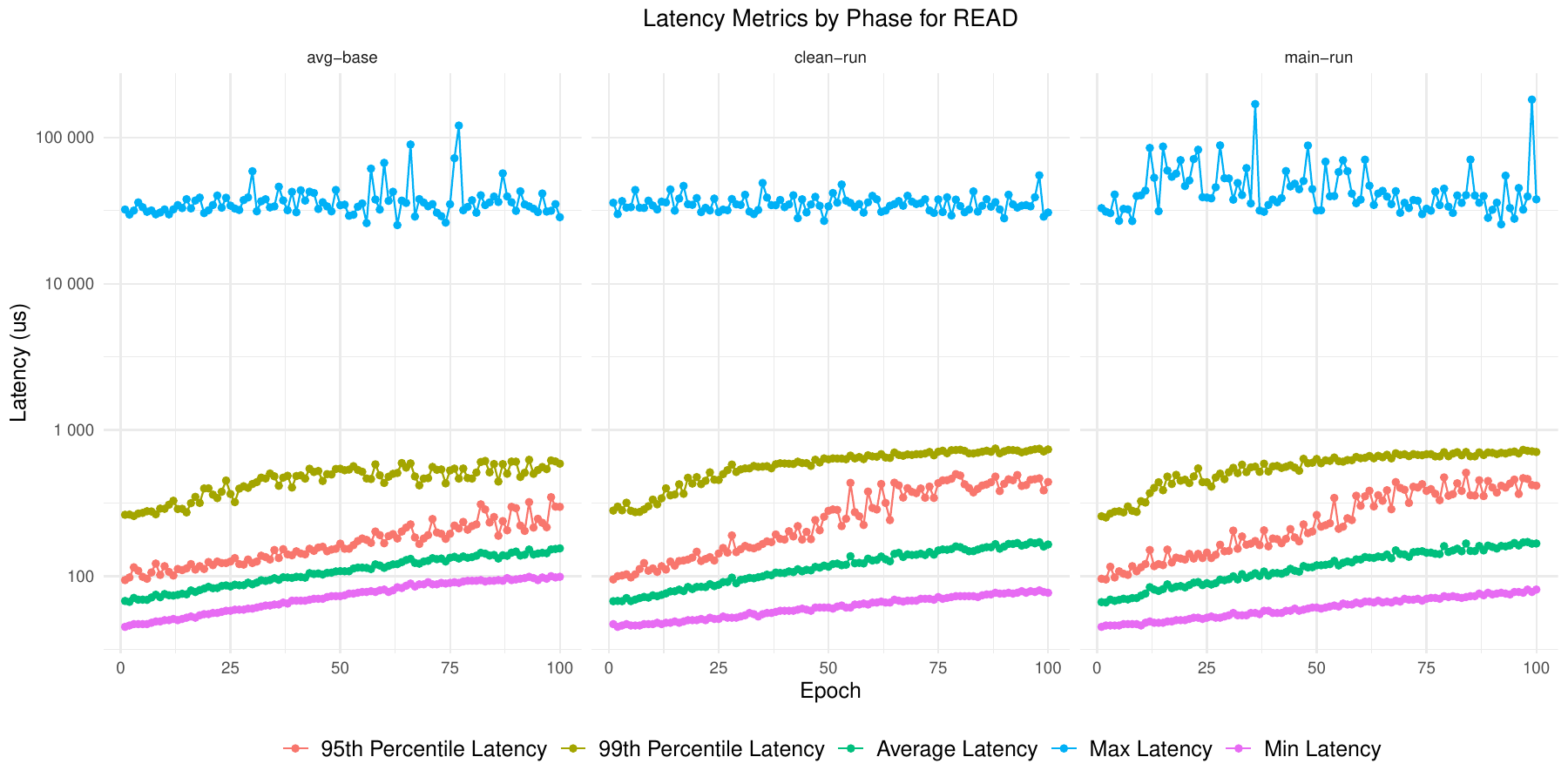}

    \caption{Behavior of latency summary statistics for \mongodb \emph{run} phases for YCSB workload C with Zipfian extensions.}
    \label{fig:mongo_latency_values}
\end{figure}

The baselines for the other platforms similarly confirm that record size is an important influence on system performance. Long records generally slow performance compared to short ones when data volume is the same. Thus record length is worth examining when benchmarking.

\subsection{Consistent spread of latencies}

We applied YCSB-IVS to \mongodb. Here we found the platform seems well able to handle the situation we are examining, where some documents are growing in length. Overall, the results for \mainrun and \cleanrun are very close, and not much worse than \avgrun: through the epochs, latency grows and throughput drops, based largely on the impact from greater data transfer because of the record size. We show, in \autoref{fig:mongo_latency_values}, charts for a variety of latency metrics on the run phases, taken in one execution of the benchmark with read-only YCSB workload C with Zipfian key distribution during the extend phase.

\begin{figure}[bth]
    \centering
    \includegraphics[width=0.7\columnwidth]{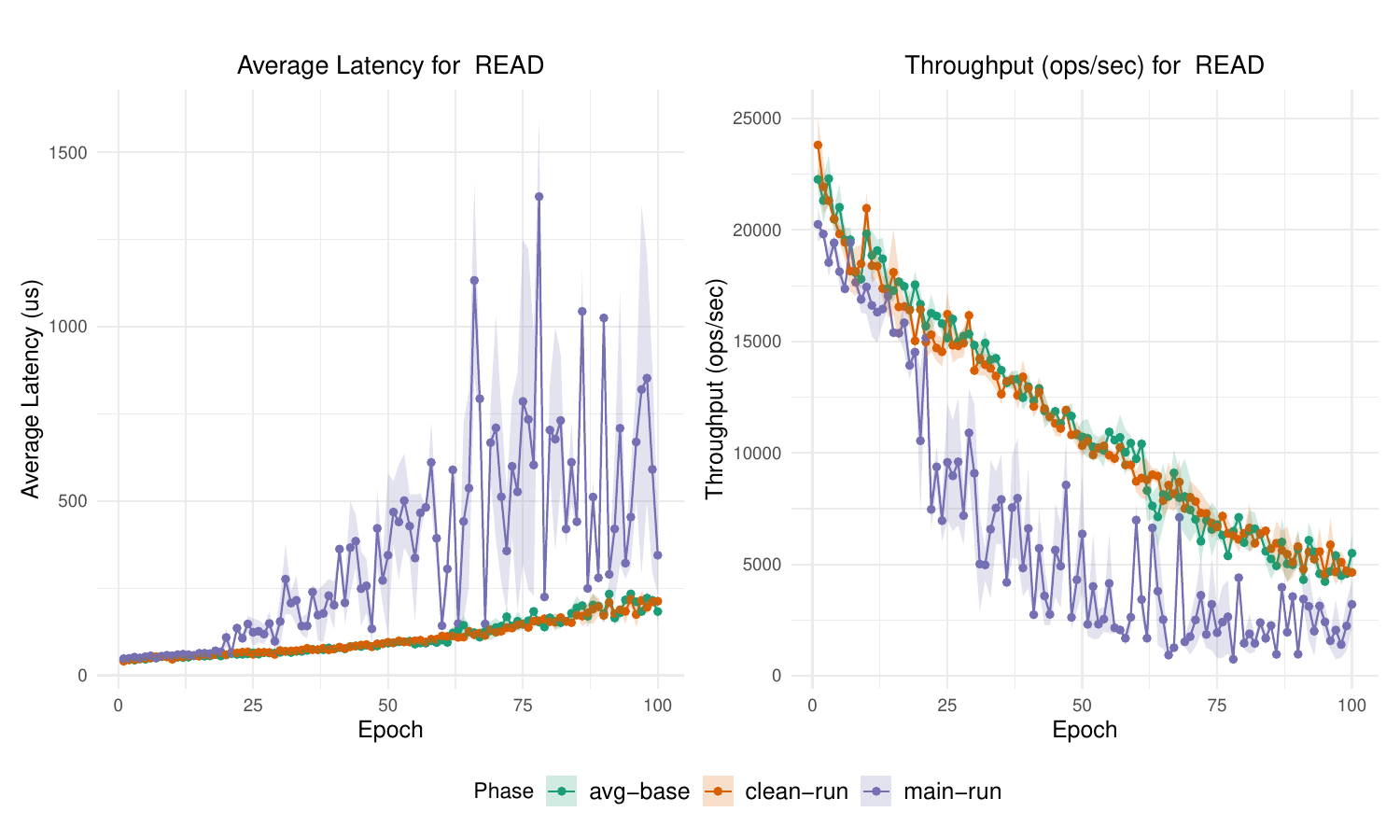}
    \caption{YCSB Workload C performance in \rocksdb with  Zipfian request distribution for Extend operations and Uniform Reads.}
    \label{fig:myrocks-heavy}
\end{figure}

\subsection{Impact of update history on query performance}
%\subsubsection*{\textbf{Heavyweight Workload.}}
%{\textbf{Heavyweight Workload.}}

%\begin{itemize}
%    \item \emph{Uniformly distributed extend operations:} As shown in \autoref{fig:myrocks-heavy}(a), latency and throughput remain similar across \mainrun, \cleanrun, and \avgrun for the first 25 extension cycles. Beyond this point, \mainrun—where queries follow data extension without reloading—exhibits slightly slower read performance than the rest of the modes. In contrast, \cleanrun; where the data is dumped and reloaded into a fresh database before querying and \avgrun; which has same sized records, maintain similar performance. However, this gap narrows over time. Overall, under uniformly performed extensions, \rocksdb indicates only a slight degradation of read efficiency on extended values over fixed-size records and the observed performance difference in \mainrun could be due to the runtime state.

%    \item \emph{Zipfian distributed extend operations:} 
%\autoref{fig:myrocks-heavy}(b) illustrates the results of extending the value sizes in each epoch using a Zipfian distribution, where a small proportion of keys were extended more frequently compared to the uniform case. In this scenario, \cleanrun exhibited only a slight deviation from \avgrun. Some overhead was noted in \mainrun in \autoref{fig:myrocks-heavy}(a), but this became much clearer in \autoref{fig:myrocks-heavy}(b) when extend operations used a Zipfian distribution. 

\autoref{fig:myrocks-heavy} shows the mean behavior of latency and throughput across five repetitions of the benchmark, with read-only YCSB workload C, and Zipfian distribution (parameter 0.99) for the Extend operations, applied to \rocksdb. We also show a measure of spread around the data points, with shading for the range: mean plus or minus 1.96 times the standard error among the five runs.
Here, the impact of a history of change is evident: results deviated significantly in \mainrun after 20 epochs. Beyond this point, the average read latency in \mainrun increased by more than an order of magnitude compared to its counterpart in \cleanrun, where a fresh copy of the data was queried. However, \cleanrun showed essentially the same performance as \avgrun, revealing that the variation in lengths between records was not impactful, only the average length of records matters.
%\end{itemize}

\subsection{Update workload with a distribution of value sizes}

\begin{figure}[htb]
    \centering
    %\begin{subfigure}{\columnwidth}
    %    \centering
        \includegraphics[width=0.9\columnwidth]{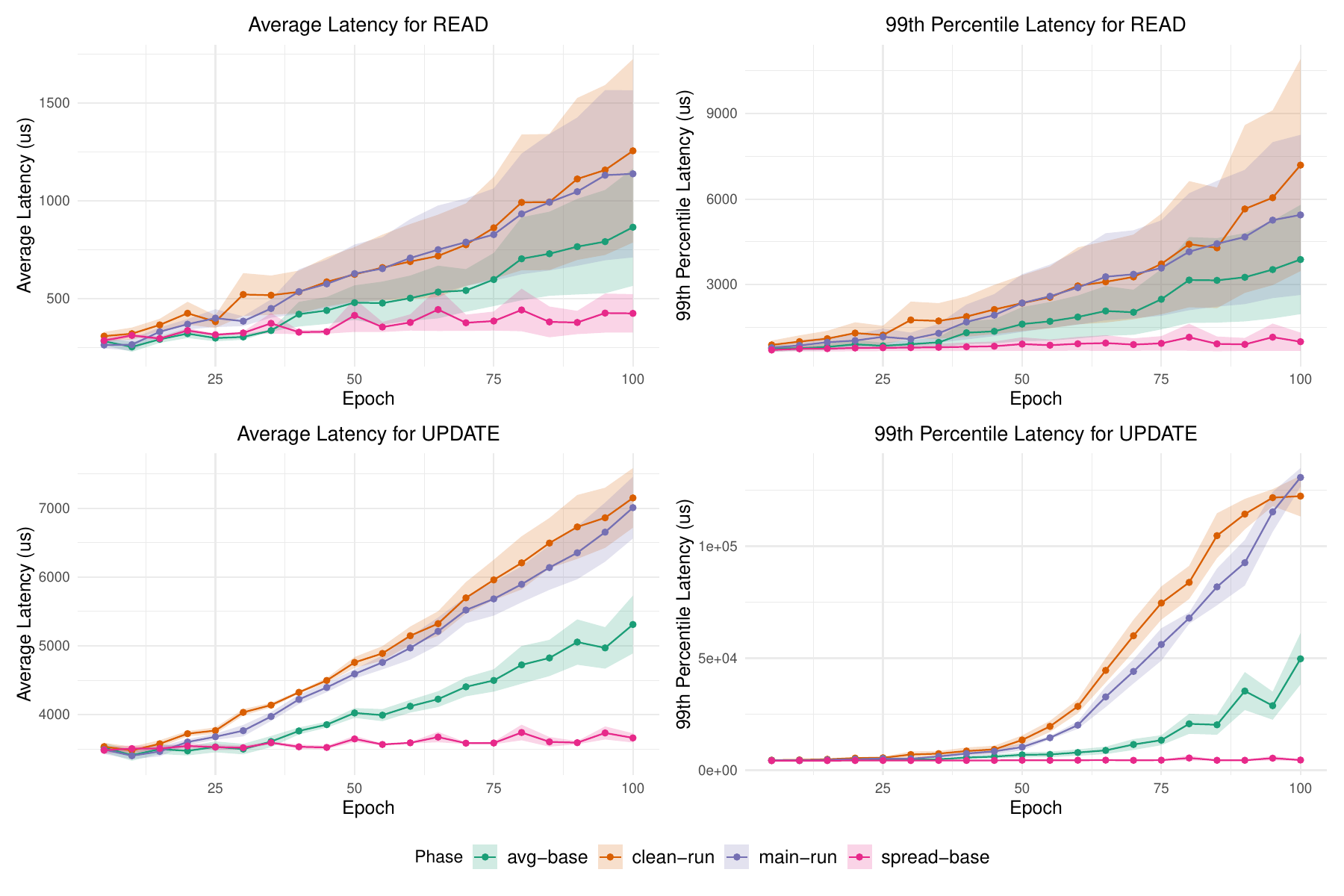}
        \caption{YCSB Workload A in \rocksdb with uniform extensions. Average and 99th percentile latency for Read and Update operations over 5 trials.}
    %\end{subfigure}
    %\vspace{0.5em} % Add spacing between the two figures
    %\hfill
    %\begin{subfigure}{\columnwidth}
    %    \centering
    %    \includegraphics[width=0.7\columnwidth]{figures/rocksdb_heavy_uniform_workloada_throughput.pdf}
    %    \caption{Average throughput over 5 trials.}
    %\end{subfigure}
    %\caption{YCSB Workload A in \rocksdb with uniform extensions}
    \label{fig:rocksdb-heavy-uniform-workloadA}
\end{figure}

We further illustrate performance under YCSB workload A that includes 50\% update operations. \autoref{fig:rocksdb-heavy-uniform-workloadA} presents results for this workload comprising 50\% reads and 50\% updates, with Extend phase using uniform distribution of keys. Note that latency is shown separately for each operation type.
% To highlight performance differences relative to baselines, we report  average behavior of mean latency, 99\textsuperscript{th} percentile latency (\autoref{fig:rocksdb-heavy-uniform-workloadA}(a)) and the average throughput (\autoref{fig:rocksdb-heavy-uniform-workloadA}(b)) across several repetitions of the workload.
%\subsubsection{\textbf{Impact of Value-size Increments:}} 

%\textbf{Value-size Increments in \rocksdb:}
Both \mainrun and \cleanrun are noticeably slower for reads, and even more so for updates (especially in the tail of the distribution), compared to \avgrun, showing that variation in length between data items can be impactful.  The spread of results from separate runs (shown by shading around the data points)  further reveal that read operations, particularly in \mainrun and \cleanrun, show greater variance in latency than updates.
%Unlike the read-only workload, which shows only minor performance overhead, the mixed workload reveals a clear decline in both read and update efficiency for \mainrun and \cleanrun compared to the baselines. Towards reaching the 100\textsuperscript{th} epoch in our experiments, updates have depicted more than 2 times 99\textsuperscript{th} percentile latency. 

Although \mainrun and \cleanrun perform similarly, \cleanrun consistently shows slightly lower efficiency. This is surprising, with a freshly populated database less effective than one built over time. %Despite this variability, the overall trend remains consistent: value-size growth, even when random, has a substantial negative impact on both read and update performance.

\subsection{Accessing highly skewed value sizes}

%\textbf{Value Size Increase in \innodb:} 

\begin{figure}[tbh!]
    \centering
    \begin{subfigure}{\columnwidth}
        \centering
        \includegraphics[width=0.9\columnwidth]{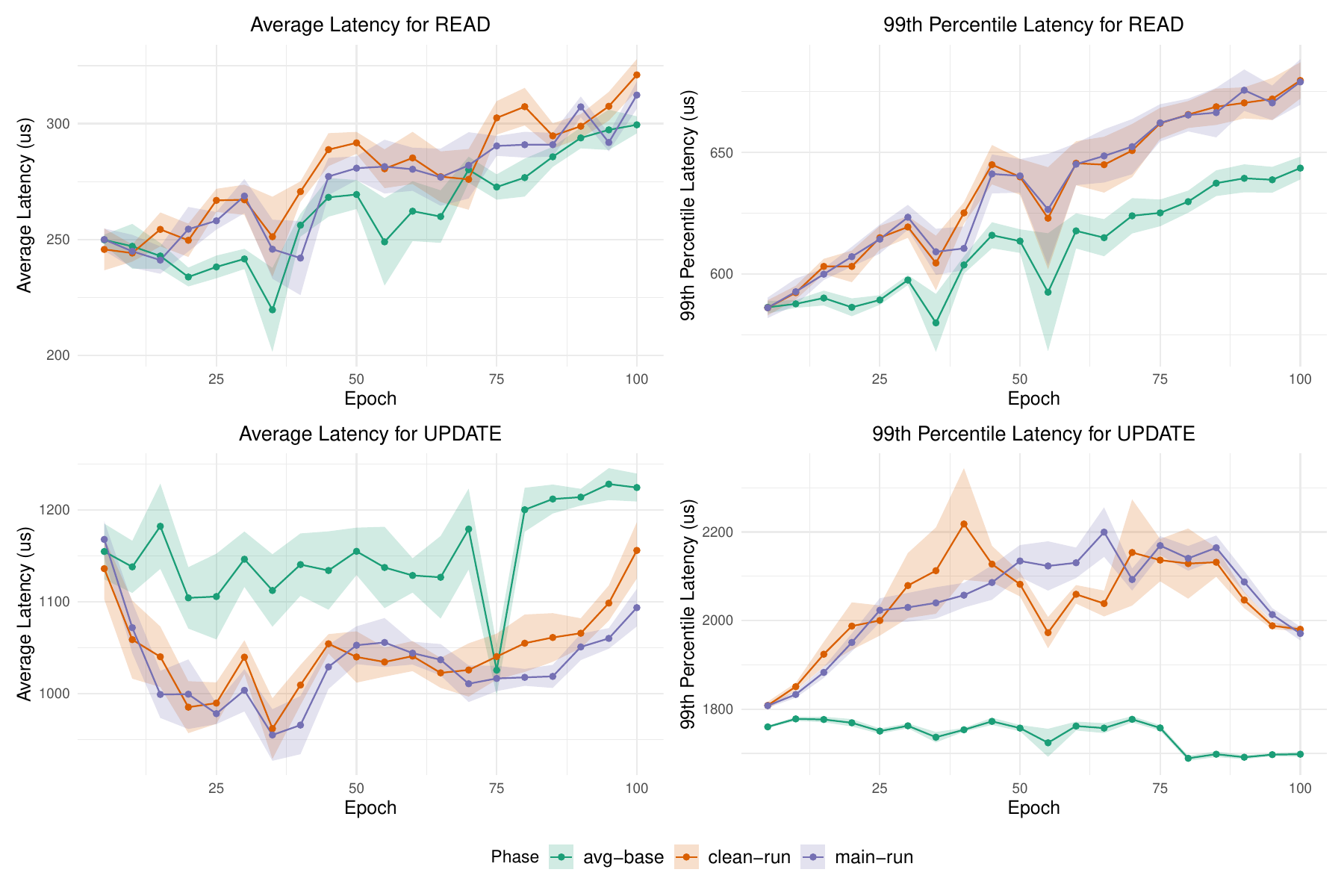}
        \caption{Average and 99th percentile latency for Read and Update operations.}
    \end{subfigure}
    %\vspace{0.5em} % Add spacing between the two figures
    \begin{subfigure}{\columnwidth}
        \centering
        \includegraphics[width=0.6\columnwidth]{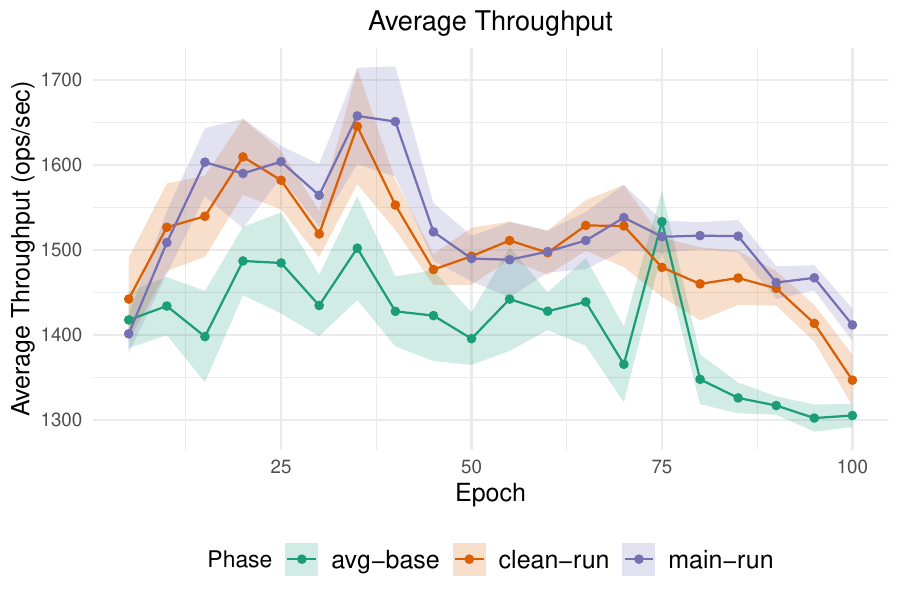}
        \caption{Average throughput.}
    \end{subfigure}
    \caption{YCSB workload A in \innodb with Zipfian extensions.}
    \label{fig:innodb-heavy-zipfian-workloadA}
\end{figure}

We conducted multiple trials on \innodb using YCSB workload A, as extend phases progressively increased value sizes with a Zipfian distribution for keys. \autoref{fig:innodb-heavy-zipfian-workloadA} presents the corresponding performance metrics.
%using the same format as \autoref{fig:rocksdb-heavy-uniform-workloadA}.
While the data seems very noisy, it does reveal that in most epochs, both \mainrun and \cleanrun perform similarly for average latency of reads and notably faster for updates, although with serious slowdown for tail latency. The throughput is mostly a bit better for \mainrun (with history) than \cleanrun (without history of extension), but both are generally higher than the \avgrun baseline.

\begin{comment}
    For this mixed workload, we observe reduced read efficiency in both \mainrun and \cleanrun compared to \avgrun in ~\autoref{fig:innodb-heavy-zipfian-workloadA}(a), consistent with earlier findings. A minor performance gap is also noticeable between \mainrun and \cleanrun for read operations.

Interestingly, for update operations, both \mainrun and \cleanrun exhibit lower average latency than \avgrun, indicating better efficiency on average. However, \avgrun demonstrates more stable tail latency, suggesting improved performance consistency under heavier load.

Overall, the average throughput results in \autoref{fig:innodb-heavy-zipfian-workloadA}(b) indicate a decline in performance for \mainrun and \cleanrun relative to the baseline, reaffirming that dynamic value-size increases adversely affected query execution across operation types.
\end{comment}

\begin{comment}
%\subsubsection{\textbf{Workload Scalability:}} 
{\textbf{Workload Scalability:}} 
The results in ~\autoref{fig:rocksdb-heavy-uniform-workloadA} and ~\autoref{fig:innodb-heavy-zipfian-workloadA} demonstrate the applicability of the proposed benchmark to arbitrary combinations of database operations, as supported by YCSB. In our experiments, we monitored the distribution of value lengths before and after each workload run and verified that it is preserved—even under workloads with frequent updates—confirming the benchmark's robustness and scalability.
\end{comment}

%% file: 8-conclusion.tex
\section{Conclusion}

We introduced a novel approach to capture the performance impact when some records become very long, distinguishing the influence of overall record length, variation in record length between items, and historic growth in length. We utilize the YCSB framework, but make substantial changes to the way runs are arranged, with four different modes of run, and a succession of epochs, showing performance under different length distributions.

Our benchmark, YCSB-IVS, has been applied to a variety of platforms. The code and all experiment data are available. 
%We have shown the impacts of our workload on document stores and relational databases, with either tree or LSM-based storage structures. using throughput and average latency as the performance measures. In general, we observe that the growth in value sizes of a specific subset of keys poses challenges that the existing benchmarks fail to uncover.
Given the presence of such a pattern of growth in practical applications, using YCSB-IVS would not only help in making a wise choice for data management, but it will also aid platform implementers to cater to such use cases.

%% file: _main.bbl
\begin{thebibliography}{10}
\providecommand{\url}[1]{\texttt{#1}}
\providecommand{\urlprefix}{URL }
\providecommand{\doi}[1]{https://doi.org/#1}

\bibitem{armstrong2013linkbench}
Armstrong, T.G., Ponnekanti, V., Borthakur, D., Callaghan, M.: Linkbench: a database benchmark based on the facebook social graph. In: Proceedings of the 2013 ACM SIGMOD International Conference on Management of Data. p. 1185–1196. SIGMOD '13, Association for Computing Machinery, New York, NY, USA (2013). \doi{10.1145/2463676.2465296}, \url{https://doi.org/10.1145/2463676.2465296}

\bibitem{battle2020database}
Battle, L., Eichmann, P., Angelini, M., Catarci, T., Santucci, G., Zheng, Y., Binnig, C., Fekete, J.D., Moritz, D.: Database benchmarking for supporting real-time interactive querying of large data. In: Proceedings of the 2020 ACM SIGMOD International Conference on Management of Data. p. 1571–1587. SIGMOD '20, Association for Computing Machinery, New York, NY, USA (2020). \doi{10.1145/3318464.3389732}, \url{https://doi.org/10.1145/3318464.3389732}

\bibitem{BensonBBLLPRSST24}
Benson, L., Binnig, C., Bodensohn, J.M., Lorenzi, F., Luo, J., Porobic, D., Rabl, T., Sanghi, A., Sears, R., T\"{o}z\"{u}n, P., Ziegler, T.: Surprise benchmarking: The why, what, and how. In: Proceedings of the Tenth International Workshop on Testing Database Systems. p. 1–8. DBTest '24, Association for Computing Machinery, New York, NY, USA (2024). \doi{10.1145/3662165.3662763}, \url{https://doi.org/10.1145/3662165.3662763}

\bibitem{biliris1992performance}
Biliris, A.: The performance of three database storage structures for managing large objects. ACM SIGMOD Record  \textbf{21}(2),  276--285 (1992)

\bibitem{bitton1983benchmarking}
Bitton, D., DeWitt, D.J., Turbyfill, C.: Benchmarking database systems-a systematic approach. Tech. rep., University of Wisconsin-Madison Department of Computer Sciences (1983)

\bibitem{cao2020characterizing}
Cao, Z., Dong, S., Vemuri, S., Du, D.H.: Characterizing, modeling, and benchmarking $\{$RocksDB$\}$$\{$Key-Value$\}$ workloads at facebook. In: 18th USENIX Conference on File and Storage Technologies (FAST 20). pp. 209--223. USENIX Association (2020)

\bibitem{Carey1986}
Carey, M.J., DeWitt, D.J., Richardson, J.E., Shekita, E.J.: Object and file management in the exodus extensible database system. In: Proceedings of the 12th International Conference on Very Large Data Bases. p. 91–100. VLDB '86, Morgan Kaufmann Publishers Inc., San Francisco, CA, USA (1986)

\bibitem{cheng2022taobench}
Cheng, A., Shi, X., Kabcenell, A., Lawande, S., Qadeer, H., Chan, J., Tin, H., Zhao, R., Bailis, P., Balakrishnan, M., et~al.: Taobench: An end-to-end benchmark for social network workloads. Proceedings of the VLDB Endowment  \textbf{15}(9),  1965--1977 (2022)

\bibitem{ycsb}
Cooper, B.F., Silberstein, A., Tam, E., Ramakrishnan, R., Sears, R.: Benchmarking cloud serving systems with ycsb. In: Proceedings of the 1st ACM Symposium on Cloud Computing. p. 143–154. SoCC '10, Association for Computing Machinery, New York, NY, USA (2010). \doi{10.1145/1807128.1807152}, \url{https://doi.org/10.1145/1807128.1807152}

\bibitem{DeyFNR14}
Dey, A., Fekete, A.D., Nambiar, R., R{\"{o}}hm, U.: {YCSB+T:} benchmarking web-scale transactional databases. In: Workshops Proceedings of the 30th International Conference on Data Engineering Workshops, {ICDE} 2014, Chicago, IL, USA, March 31 - April 4, 2014. pp. 223--230. {IEEE} Computer Society (2014). \doi{10.1109/ICDEW.2014.6818330}, \url{https://doi.org/10.1109/ICDEW.2014.6818330}

\bibitem{du1992query}
Du, W., Krishnamurthy, R., Shan, M.C.: Query optimization in a heterogeneous dbms. In: VLDB. vol.~92, pp. 277--291 (1992)

\bibitem{ErlingALCGPPB15}
Erling, O., Averbuch, A., Larriba-Pey, J., Chafi, H., Gubichev, A., Prat, A., Pham, M.D., Boncz, P.: The ldbc social network benchmark: Interactive workload. In: Proceedings of the 2015 ACM SIGMOD International Conference on Management of Data. p. 619–630. SIGMOD '15, Association for Computing Machinery, New York, NY, USA (2015). \doi{10.1145/2723372.2742786}, \url{https://doi.org/10.1145/2723372.2742786}

\bibitem{funke2011benchmarking}
Funke, F., Kemper, A., Neumann, T.: Benchmarking hybrid oltp\&olap database systems. In: Datenbanksysteme für Business, Technologie und Web (BTW), pp. 390--409. Gesellschaft f{\"u}r Informatik eV, Bonn (2011)

\bibitem{gardarin1996calibrating}
Gardarin, G., Sha, F., Tang, Z.H.: Calibrating the query optimizer cost model of iro-db, an object-oriented federated database system. In: VLDB. vol.~96, pp.~3--6. Citeseer (1996)

\bibitem{gray1993database}
Gray, J.: Database and transaction processing performance handbook. (1993)

\bibitem{gupta2020smartbench}
Gupta, P., Carey, M.J., Mehrotra, S., Yus, o.: Smartbench: A benchmark for data management in smart spaces. Proceedings of the VLDB Endowment  \textbf{13}(12),  1807--1820 (2020)

\bibitem{hurson1993object}
Hurson, A.R., Pakzad, S.H., Cheng, J.B.: Object-oriented database management systems: evolution and performance issues. Computer  \textbf{26}(2),  48--58 (1993)

\bibitem{idreos2020key}
Idreos, S., Callaghan, M.: Key-value storage engines. In: Proceedings of the 2020 ACM SIGMOD International Conference on Management of Data. p. 2667–2672. SIGMOD '20, Association for Computing Machinery, New York, NY, USA (2020). \doi{10.1145/3318464.3383133}, \url{https://doi.org/10.1145/3318464.3383133}

\bibitem{iyer1994data}
Iyer, B.R., Wilhite, D.: Data compression support in databases. In: VLDB. vol.~94, pp. 695--704 (1994)

\bibitem{jain2021dgraph}
Jain, M.: Dgraph: Synchronously replicated,, transactional and distributed graph database (2021), \url{https://github.com/hypermodeinc/dgraph/blob/master/paper/dgraph.pdf}

\bibitem{KimHYK23}
Kim, S., Hoang, Y., Yu, T.T., Kanwar, Y.S.: Geoycsb: {A} benchmark framework for the performance and scalability evaluation of geospatial nosql databases. Big Data Res.  \textbf{31},  100368 (2023). \doi{10.1016/J.BDR.2023.100368}, \url{https://doi.org/10.1016/j.bdr.2023.100368}

\bibitem{myrocks_vars}
{MariaDB}: {MariaDB: MyRocks System Variables} (2025), \url{https://mariadb.com/kb/en/myrocks-system-variables/}, accessed: 2025-05-30

\bibitem{mongodb_vars}
{MongoDB}: {MongoDB Server Parameters for a Self-Managed Deployment} (2025), \url{https://www.mongodb.com/docs/manual/reference/parameters/}, accessed: 2025-06-02

\bibitem{o2009star}
O'Neil, P., O'Neil, E., Chen, X., Revilak, S.: The star schema benchmark and augmented fact table indexing. In: Nambiar, R., Poess, M. (eds.) Performance Evaluation and Benchmarking. pp. 237--252. Springer Berlin Heidelberg, Berlin, Heidelberg (2009)

\bibitem{pelekis2004literature}
Pelekis, N., Theodoulidis, B., Kopanakis, I., Theodoridis, Y.: Literature review of spatio-temporal database models. The Knowledge Engineering Review  \textbf{19}(3),  235--274 (2004)

\bibitem{ray2011jackpine}
Ray, S., Simion, B., Demke~Brown, A.: Jackpine: A benchmark to evaluate spatial database performance. In: 2011 IEEE 27th International Conference on Data Engineering. pp. 1139--1150. IEEE (2011). \doi{10.1109/ICDE.2011.5767929}

\bibitem{shapiro1999managing}
Shapiro, M., Miller, E.: Managing databases with binary large objects. In: 16th IEEE Symposium on Mass Storage Systems in cooperation with the 7th NASA Goddard Conference on Mass Storage Systems and Technologies (Cat. No.99CB37098). pp. 185--193. IEEE (1999). \doi{10.1109/MASS.1999.830036}

\bibitem{TheodorakisCW24}
Theodorakis, G., Clarkson, J., Webber, J.: An empirical evaluation of variable-length record b+trees on a modern graph database system. In: 40th International Conference on Data Engineering, {ICDE} 2024 - Workshops, Utrecht, Netherlands, May 13-16, 2024. pp. 343--349. {IEEE} (2024). \doi{10.1109/ICDEW61823.2024.00051}, \url{https://doi.org/10.1109/ICDEW61823.2024.00051}

\bibitem{TPC}
(TPC), T.P.P.C.: Tpc homepage. \url{https://www.tpc.org/} (2024), accessed: 2024-12-20

\bibitem{wang2008measurement}
Wang, X., Schulzrinne, H., Kandlur, D., Verma, D.: Measurement and analysis of ldap performance. IEEE/ACM Transactions On Networking  \textbf{16}(1),  232--243 (2008)

\bibitem{westmann2000implementation}
Westmann, T., Kossmann, D., Helmer, S., Moerkotte, G.: The implementation and performance of compressed databases. ACM Sigmod Record  \textbf{29}(3),  55--67 (2000)

\end{thebibliography}
